\documentclass[aps,prb,twocolumn,superscriptaddress,floatfix]{revtex4}
\usepackage{graphicx}
\usepackage{psfrag}
\usepackage{color}
\usepackage{subfigure}
\usepackage{amsmath}
\definecolor{dred}{rgb}{0.7,0.0,0.0}
\allowdisplaybreaks 
 
\begin{document}

\title{Charge Stripes in the Two-Orbital Hubbard Model for Pnictides}

\author{Qinlong Luo}
 
\affiliation{Department of Physics and Astronomy, The University of
  Tennessee, Knoxville, TN 37996} 
\affiliation{Materials Science and Technology Division, Oak Ridge
  National Laboratory, Oak Ridge, TN 32831} 

\author{Dao-Xin Yao}
\affiliation{State Key Lab of Optoelectronic Materials and Technologies, School of
Physics and Engineering, Sun Yat-sen University, Guangzhou 510275, China}

\author{Adriana Moreo}
\author{Elbio Dagotto}

\affiliation{Department of Physics and Astronomy, The University of
  Tennessee, Knoxville, TN 37996} 
\affiliation{Materials Science and Technology Division, Oak Ridge
  National Laboratory, Oak Ridge, TN 32831}

\date{\today}
%\maketitle

\begin{abstract}
The two-orbital Hubbard model for the pnictides is studied numerically 
using the real-space Hartree-Fock approximation on finite clusters. 
Upon electron doping, states with a nonuniform distribution of charge 
are stabilized. The observed patterns correspond to charge stripes 
oriented perpendicular to the  direction of the spin stripes 
of the undoped magnetic ground state. While these  charge
striped states are robust when the undoped state has a Hubbard gap, their
existence when the intermediate-coupling magnetic metallic state of pnictides is doped 
was also observed for particular model parameters.
Results for hole doping and implications 
for recent experiments that reported electronic nematic states 
and spin incommensurability are also briefly discussed.


\end{abstract}

\maketitle

\section{Introduction}

The recent discovery of the iron-based pnictide 
superconductors\cite{johnston} and the perception that the pairing mechanism 
may be triggered by antiferromagnetic fluctuations, similarly as in the cuprates,
has motivated a considerable theoretical effort to understand the properties
of multiorbital Hubbard models. Using mean-field,
random phase approximation, and computational techniques, several groups 
have addressed the nature of the undoped ground state of Hubbard models 
and the dominant pairing mechanisms 
upon doping.\cite{theory,maria,moreo,raghu,rong,luo} 
Among the most robust results recently unveiled are {\it (i)} 
the existence of an intermediate Hubbard
$U$ coupling regime where the undoped state is simultaneously antiferromagnetic
and metallic\cite{rong,luo} as found in transport and neutron scattering 
experiments,\cite{johnston} 
and {\it (ii)} the presence of superconducting tendencies upon charge doping mainly 
in the $A_{\rm 1g}$ channel (extended $s\pm$, with nodes),\cite{theory,maria,moreo,andrew}  
with the nodal pairing states 
$B_{\rm 1g}$ and $B_{\rm 2g}$ located close in energy.\cite{maria,moreo,andrew}

While the superconductivity and $(\pi,0)$ magnetism have received most of the
attention, recent experiments have revealed an even more complex behavior in pnictides.
For example,
inelastic neutron scattering (INS) experiments reported evidence 
for spin incommensurability (spin IC) 
in the superconducting state of the 
hole very overdoped pnictide KFe$_2$As$_2$.\cite{spinIC} 
This material does not have electron-pockets in its band 
structure due to the heavy doping (50\%) and, as a consequence, nesting mechanisms cannot 
produce the spin IC, whose origin then remains puzzling. Neutron studies for 
FeSe$_{0.5}$Te$_{0.5}$~\cite{lumsden,FST} have also reported low-energy spin IC peaks 
near the spin resonance.~\cite{FST}
Electron-doped pnictides such as Ba(Fe,Co)$_2$As$_2$ do not show spin IC 
at low energies,\cite{edoped} but above 50 meV there is a splitting of 
the $(\pi,0)$ peak. 

In parallel to these developments, scanning tunneling microscopy (STM) 
experiments on underdoped Ca(Fe$_{0.97}$Co$_{0.03}$)$_2$As$_2$ reported 
the existence of static unidirectional electronic nanostructures of 
dimension 8 $a_{\rm Fe-Fe}$ (with $a_{\rm Fe-Fe}$ the inter-iron distance) 
along the $a$-axis.\cite{seamus} This electronic nematic order state is 
qualitatively similar to those widely discussed in other materials, such as in
the high-temperature superconductors based on copper.\cite{zaanen,emery} Note
that the lines of spins that point in the same direction, namely the orientation
of the $(\pi,0)$ spin stripes, is actually along the $b$-axis, namely the nematic order
and the spin-stripe order are perpendicular to one another. 

The spin IC and nematic order reveal new similarities
between the Cu- and Fe-based high temperature superconductors, more than expected
considering the widely accepted perception that the former are in the strongly correlated
regime of Hubbard $U$ couplings while the latter are at weak or intermediate coupling.
In the Cu-oxide context, both spin IC and several electronic anisotropies in transport
have been rationalized in terms of charge {\it striped} states, either static or
dynamic.\cite{emery} Could it be that similar states are also of relevance in the pnictides?

With this motivation, in the present effort the possible existence of
charge striped states in Hubbard models for the pnictides will be explored. Within the
real-space Hartree-Fock approximation,
it will be shown below that the
two-orbital Hubbard model indeed displays charge striped states upon doping.
The charge, spin, and orbital properties of these states will be discussed.
The properties of the striped states reported below are robust when the coupling
$U$ is above the critical value needed to open a gap in the undoped limit. 
Considering the complex nature of multiorbital systems,
it is important to document the properties of these striped states even in gapped
regimes that at first sight are unrelated to undoped pnictides that are known to be
(bad) metals. However,   
the charge amplitude of the stripes (namely the difference between the
largest and smallest values of the charge at every site in the striped state) decreases with decreasing gap, and
an interesting observation is that (weak) stripes are still found upon doping the
intermediate coupling magnetic and {\it metallic} state of the undoped limit for particular hopping
parameters. This result is conceptually different from the previous investigations of
stripes in doped large-$U$ Hubbard insulators.

\section{Model and Method}

In this publication, the Hubbard
model~\cite{maria,moreo,raghu} based on the $d_{xz}$ and
$d_{yz}$ Fe orbitals will be investigated. The use of these two orbitals
is reasonable since they provide the
largest contribution to the pnictide's Fermi Surface.\cite{phonon}
Studies involving models with more orbitals
are certainly important, but they will be addressed 
only in future investigations after introducing
the present novel findings 
based on two orbitals. A variety of other results that have recently been
gathered for the two-orbital model~\cite{maria,moreo,andrew} have already clearly 
shown that this model is in good agreement with
experiments in several respects, including the existence of $(\pi,0)$-$(0,\pi)$ magnetic order,
a Fermi surface that by construction is similar to those found in photoemission, and pairing
channels that include the $A_{\rm 1g}$ state (of which the 
s$\pm$ pairing is a special case) as well
as competitors with $B_{\rm 1g}$ and $B_{\rm 2g}$ symmetries.
Note also that the use of five orbital models would demand a much
larger computational effort than reported here, since the diagonalizations used in the iterative
procedure to solve the HF equations scales as $(2N)^4$, with $N$ the number of sites of the real-space cluster 
and $2$ the number of
orbitals. Thus, a similar calculation for $5$ orbitals will require $(5/2)^4$$\sim$39 times more computer time.
It is for these reasons that it is reasonable to report first 
the analysis of the case of two orbitals, identifying
the main tendencies,
and then in the near future increase the number of orbitals. In the case of five orbitals, recent
{\it ab-initio} calculations supplemented by the construction of a model Hamiltonian led to
predictions for the values of all the couplings that will be used in our model.\cite{lda-model} 
However, since not all pnictides have the same couplings, 
and since the calculation of Ref.~\onlinecite{lda-model}
involves approximations, in our present effort, and in future efforts with more orbitals, 
an exploration will be carried out varying the couplings in
a reasonable range (mainly focussing on intermediate $U$ values, 
as suggested in Refs.~\onlinecite{rong,luo}), as opposed
to fixing those couplings to a unique set.

As hopping amplitudes in the tight-binding portion of the model, 
fits to band calculations~\cite{raghu} as well
as Slater-Koster (SK) hoppings~\cite{maria,moreo} will be used.
The electronic interaction is given by the standard Hubbard term, which has an
intraorbital repulsion $U$, a Hund coupling $J_{\rm H}$, an interorbital
repulsion $U'$ (fixed as $U'$=$U$-$2J_{\rm H}$ here), 
and a pair-hopping term with coupling $J'$=$J_{\rm H}$.\cite{oles83}
The explicit model does not need to be reproduced here since it is well known from
several previous investigations.\cite{maria,moreo,raghu} 
In the standard Hartree-Fock (HF)
approximation in real space, the first term with the intraorbital repulsion $U$ becomes
$H^{\rm HF}_1$=$U~\sum_{{\bf i} \alpha \sigma \sigma'} (-1)^{1-\delta_{\sigma,\sigma'}}
[        c^\dagger_{{\bf i}, \alpha,  \sigma} c_{{\bf i}, \alpha,  \sigma'} 
 \langle c^\dagger_{{\bf i}, \alpha, -\sigma} c_{{\bf i}, \alpha, -\sigma'} \rangle
-{{1}\over{2}}  \langle c^\dagger_{{\bf i}, \alpha, \sigma} c_{{\bf i}, \alpha, \sigma'} \rangle
 \langle c^\dagger_{{\bf i}, \alpha, -\sigma} c_{{\bf i}, \alpha, -\sigma'} \rangle ]$.
The other terms involve
HF expressions too long to be reproduced here, 
but they are standard and the  reader can find the full HF Hamiltonian 
in the supplementary information.\cite{supplementary} 
The minimization of the HF energy with respect to the
various expectation values
was carried out numerically via an iterative process,\cite{numerical} mainly employing 16$\times$16 clusters,
with a starting configuration for the expectation values 
chosen at random at every site (i.e. without biasing toward striped states). Several clusters 
smaller than 16$\times$16 were also explored and  size effects
do not seem large.

%{\it Results.} 

\section{Results}

\subsection{Hoppings of Refs.~\onlinecite{maria,moreo}}

Let us start the description of our main results 
using the Slater-Koster hoppings introduced in Refs.~\onlinecite{maria,moreo}.
A wide range of electronic densities $\langle n \rangle$ were numerically
explored by varying the chemical potential $\mu$. In most cases, 
the results of the HF model energy minimization were sufficiently clear that
they admit a simple discussion, and the emphasis below will be on those 
special cases, such as $\langle n \rangle$$\sim$$2.3$. At these densities, 
several starting configurations for the unknown expectation values
that appear in the HF model were chosen via a random number 
generator and it was
observed that the
iterative process leads to nearly identical solutions or, if this was  not the case, to 
local minima solutions 
with a higher energy. When different types of solutions were found, of course only 
those with the lowest energy
were kept and analyzed.

Following the energy minimization criterion,
the charge patterns found in some other cases were slightly more complicated,
with segments of stripes clearly formed at the local level but sometimes with these
segments not properly merged together to form nearly perfectly spaced stripes. Since these
states still have all the characteristics of stripe states, it may occur that for
a particular set of couplings
some stripe configurations do not fit properly in the cluster sizes used and geometrical 
frustration effects
may lead to the partial breaking of the stripes. So while the focus below is on the most
clear cases, the other states found 
are all ``stripy'' in nature, thus our results seem generic of a broad range
of couplings and densities.\cite{lorenzana} 

\begin{figure}[thbp]
\begin{center}
\includegraphics[width=8.5cm,clip,angle=0]{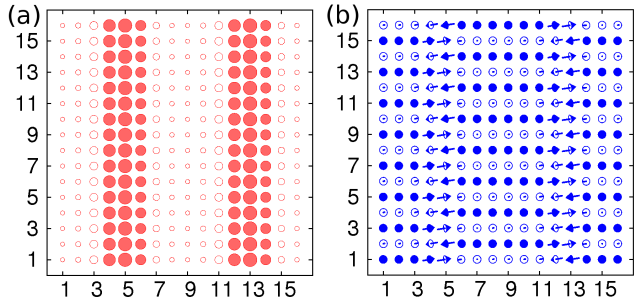}
\caption{(color online) 
(a) Example of charge striped state found in the HF approximation to the two-orbital model, using
the hoppings of Refs.~\onlinecite{maria,moreo}, $\langle n \rangle$=2.33, 
$U$=0.8, $J_{\rm H}$/$U$=0.25, and a 16$\times$16 cluster. 
The size of the circles is linearly related to  the charge, 
with the largest circles denoting $n_{\rm max}$=2.42 and the smallest $n_{\rm min}$=2.27. 
Here, and in the other figures,
full (open) circles are used when the local 
density is larger (smaller) than the average.
(b) Mean value of the spin in the state shown in (a). Note the presence of domain walls at the location 
of the charge stripes, inserted in a mainly $(0,\pi)$ background.
}
%\vskip -0.6cm
\label{Fig1}
\end{center}
\end{figure}

Figure~\ref{Fig1}~(a) contains a typical HF state that 
has been observed repeatedly in our studies, even using several different
starting configurations for the iterations. It is clear that the charge is not
uniformly distributed but it forms vertical stripes, breaking spontaneously 
rotational invariance. 
Of course, the $\pi/2$-rotated state is also a degenerate solution (horizontal stripes)
and the convergence to one
or the other depends on the randomly chosen initial state for
the iterative process. Note that these stripes 
are ``weaker'' than the usual Cu-oxides HF stripes
in the sense that the charge difference $\Delta n$ 
 ($\sim$0.15 in Fig.~\ref{Fig1}~(a)) 
between the maximum $n_{\rm max}$ and minimum $n_{\rm min}$ local 
charges is not as large as in the cuprates where $\Delta n$ is of order 1. 
This simply 
arises from the values of the coupling $U$ that were investigated that are smaller for pnictides 
than cuprates, in units of the bandwidth. 
Also it is interesting to note that $n_{\rm min}$ in Fig.~\ref{Fig1}~(a)
is 2.27. Thus, the regions in the striped state that have the less charge still deviate from 
the ``undoped'' limit, and they are 
electron doped as the rest of the striped state, while in the cuprates 
the regions between the hole stripes have densities very close to those of the
undoped insulator. From this perspective, our striped states should be metallic, 
in agreement with the density-of-states (not shown) that has
a nonzero weight at the chemical potential $\mu$.

The expectation values of the spin at each site, defined as  
${\vec s}_{\bf i}$=$\langle \sum_{\alpha,\sigma,\sigma'} c^\dagger_{{\bf i},\alpha,\sigma}
{\vec \sigma}_{\sigma \sigma'} c_{{\bf i},\alpha,\sigma'} \rangle$,
that correspond to the striped state shown in Fig.~\ref{Fig1}~(a) 
are given in
Fig.~\ref{Fig1}~(b). The spin pattern is dominated by the $(0,\pi)$ configuration, with spins
parallel along one axis and antiparallel along the other. However, at the locations of the maximum electronic
densities in Fig.~\ref{Fig1}~(a), the $(0,\pi)$ spin order breaks locally  
and domain walls are formed,\cite{walls} 
with the spins where the local charge is  maximized 
presenting an orientation nearly perpendicular to the rest.
In a HF minimization problem, where all the expectation values are entangled, 
it is difficult to establish if the
spin state with walls drives the charge stripes or vice versa. 
But by analogy with the cuprates, it can be expected
that the spin state $(0,\pi)$ tries to expel the extra charge since it is disruptive to such a state, and 
such excess charge is located at the walls where the
spin does not maintain locally the $(0,\pi)$ order. Of course, all these statements
made for vertical stripes and spin $(0,\pi)$ order are the same for the rotated degenerate
configuration with horizontal stripes and $(\pi,0)$ spin order.

The HF state Figs.~\ref{Fig1}~(a,b) has other interesting properties: 
{\it (i)} The orbital $d_{yz}$ is more populated than $d_{xz}$ in regions 
with the lowest local charge, i.e. where the spin order is locally
$\sim$$(0,\pi)$. In our opinion, this should not be considered as
indicative of long-range orbital order but instead it indicates an orbital weight 
``redistribution'' induced by the spin order that breaks rotational invariance, as recently
discussed elsewhere.\cite{owr} 
Where the local charge is the largest, on the other hand, both orbitals 
are populated approximately equally. 
{\it (ii)} The deviations from a perfect $(0,\pi)$ spin state
induce a small shift away from $(0,\pi)$ in the 
spin structure factor 
%$S({\bf k})$ 
peak position, towards $(\pi,\pi)$. 
Then, the HF striped states can produce spin incommensurability, as found in some
experiments (see Introduction).
{\it (iii)} Note that the charge stripes and the spin stripes (i.e. the lines of parallel 
spins in the $(0,\pi)$ arrangement) are {\it perpendicular} to one another, 
as observed in the STM experiments where nematic order was
reported.\cite{seamus}

\begin{figure}[thbp]
\begin{center}
\includegraphics[width=8.5cm,clip,angle=0]{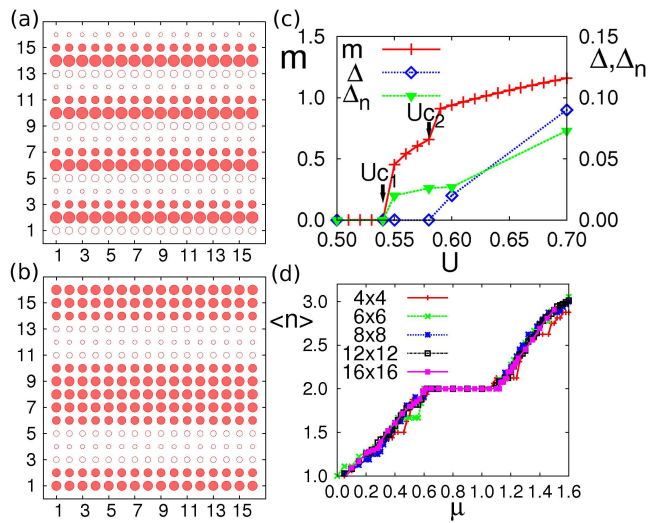}
\caption{(color online)  
(a) HF charge striped state using
the hopping amplitudes of Refs.~\onlinecite{maria,moreo}, at
$\langle n \rangle$=2.45, $U$=0.8, and $J_{\rm H}$/$U$=0.25. 
The size of the circles is proportional to the charge, 
with $n_{\rm max}$=2.48 and $n_{\rm min}$=2.41.
(b) Same as (a) but for $\langle n \rangle$=1.83, $U$=1.0, with 
$n_{\rm max}$=1.84 and $n_{\rm min}$=1.81.
(c) $(\pi,0)$ antiferromagnetic order parameter $m$, charge
gap $\Delta$ (from $\langle n \rangle$ vs. $\mu$), and 
$\Delta n$=$n_{\rm max}$-$n_{\rm min}$ at $\langle n \rangle$$\sim$2.3, as a function of $U$.
(d) $\langle n \rangle$ vs. $\mu$ at $U$=1.0, $J_{\rm H}$/$U$=0.25, 
and various lattice
sizes, suggesting that size effects are small  in this quantity. Results
at others $U$'s appear equally well converged.
}
%\vskip -0.6cm
\label{Fig2}
\end{center}
\end{figure}

Figure~\ref{Fig2}~(a) contains another typical
HF state with charge stripes that was found in our studies, 
at an electronic density $\langle n \rangle$ larger than in Fig.~\ref{Fig1}~(a), thus
concomitantly inducing a reduction in the distance between stripes. It is interesting to observe that
$\Delta$$n$ is smaller to 
that in Fig.~\ref{Fig1}~(a), 
suggesting that there must be an ``optimal'' doping for vertical/horizontal 
stripe formation, as found in cuprate's investigations, 
since in the undoped case $\langle n \rangle$=2.0 there are no stripes and thus $\Delta$$n$=0. 
Also, note that our results even farther from $\langle n \rangle$=2.0 indicate a variety of complex
patterns, with checkerboards, diagonal stripes, and other arrangements. However, 
since their associated spin orders are very different from the $(\pi,0)$ (or $(0,\pi))$ 
state prevailing in pnictides, and the doping is too large to be compared with
available experimental results, those exotic states will not be described here. Presumably
when the electronic density deviates substantially from the undoped case, then our
results reach a regime not yet found experimentally in these materials, or the two-orbital
model breaks down.

Figure~\ref{Fig2}~(b) contains results 
for the case of hole doping, and here the stripes are found to be weaker, with $\Delta n$$\sim$0.03. 
At least for the hopping amplitudes of Refs.~\onlinecite{maria,moreo}, 
the stripe tendencies appear stronger for electron 
doping than for hole doping. Also, in the hole-doped case the lines of parallel spins tend 
to run parallel to the charge stripes, 
instead of perpendicular as in the electron-doped case. This parallel vs. perpendicular
relative patterns of charge and spin may depend on details and may not be universal, thus
they need to be investigated in more details in the future.

Figure~\ref{Fig2}~(c) displays $\Delta n$ vs. $U$, 
at $\langle n \rangle$$\sim$2.3, 
compared with results for the undoped limit, namely 
the $(\pi,0)$ (or $(0,\pi)$) magnetic order parameter $m$ and
the charge gap $\Delta$. While $m$ rapidly grows with $U$ at $U_{\rm c1}$, 
$\Delta$ remains zero in a narrow region
beyond $U_{\rm c1}$ and only after reaching a second critical value $U_{\rm c2}$ 
it slowly increases 
with increasing $U$. This intermediate 
region is both metallic and magnetic, and has been emphasized before
as the regime of interest for pnictides.\cite{rong,luo} 
$\Delta n$ follows $\Delta$ 
with decreasing $U$ from strong coupling but it appears
to survive, albeit with a small value, in the intermediate magnetic/metallic phase. 
If this result would survive the introduction 
of fluctuations beyond the HF approximation,  
the undoped metallic magnetic state
would admit weak charge stripes upon 
doping, a conceptually novel result since
previous stripe efforts have focused on their presence only when 
doping a Hubbard insulator. For those  insulators the rationale for the presence
of stripes was dominated by the fact that extra charges tend to disrupt the
regular spin order of the undoped case. Thus, to reduce this ``damage'' to spin
order, the extra charge is arranged forming stripes. Such a way to understand the
stripes of the cuprates can be extended to pnictides even in the magnetic and
metallic regime: the extra charge may need to be accumulated in patterns,
as opposed to regularly distributed, to minimize the energy damage to the spin order.
Note also that size effects do not seem large here
since some quantities are already converged on the 16$\times$16 cluster 
(see Fig.~\ref{Fig2}~(d)).
% for $\langle n \rangle$ vs. $\mu$).

\begin{figure}[thbp]
\begin{center}
\includegraphics[width=8.5cm,clip,angle=0]{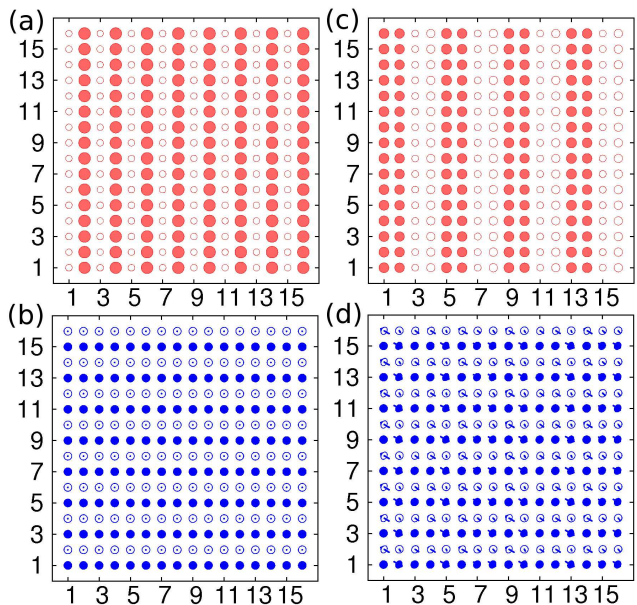}
\caption{(color online)  
(a) Charge striped state solution of the HF two-orbital Hubbard model, using
the hoppings of Ref.~\onlinecite{raghu}, $\langle n \rangle$=2.36, 
$U$=0.50, and $J_{\rm H}$/$U$=0.25. 
The size of the circles is proportional to the charge, 
with $n_{\rm max}$=2.365 and $n_{\rm min}$=2.361 (a very weak charge stripe). 
(b) Spin state associated with (a).
(c) Same as (a) but for $U$=0.90, $\langle n \rangle$=2.14, 
$n_{\rm max}$=2.141, and $n_{\rm min}$=2.139
(d) Spin state associated with (c).
}
%\vskip -0.6cm
\label{Fig3}
\end{center}
\end{figure}

\subsection{Hoppings of Ref.~\onlinecite{raghu}}

The HF results obtained using another set of hopping amplitudes
(Ref.~\onlinecite{raghu}) are presented in Fig.~\ref{Fig3}. In this
case, the analysis was slightly more involved, since 
%several overall
%densities $\langle n \rangle$ could not be stabilized by varying $\mu$, and, in addition,
the convergence of the HF energy minimization was slower, 
with the iterative process for
convergence sometimes ending in metastable states with charge inhomogeneous irregular patterns
(but still displaying stripes locally).
%
%that sometimes resembled stripes but sometimes checkerboards or other arrangements. 
%For this reason, here only a few clear cases of charge stripes will be presented. 
However, in many  cases the convergence to a nearly perfect stripe pattern 
was clear.
Panels Figs.~\ref{Fig3}(a) and (c) display the charge order in the
HF states at two different 
$\langle n \rangle$'s and two $U$'s with a magnetic 
undoped state (the hopping unit $t_1$ of Ref.~\onlinecite{raghu}
is taken as 0.2~eV here, as in Ref.~\onlinecite{andrew}).  
In both cases, charge stripes can be identified,
with $\Delta$$n$ growing with  $U$, as in Fig.~\ref{Fig2}(c). At
$U$=0.50 the charge gap is negligible, so these (weak) charge stripes
appear in the metallic magnetic state.
%was found, but it is small 
%(0.2 at $U$=0.50 and $\sim$0.03 at $U$=0.32).
The associated spin patterns also 
present a striped arrangement, and the spin and charge
stripes are perpendicular. 
%Recalling the
%observation of both arrangements also in the previously studied hoppings set,
%it appears that details can alter the balance between the two relative
%orientations. 
There are some differences between the stripes
of Figs.\ref{Fig1} and \ref{Fig3}, such as the stripe periodicity.
These differences are to be expected since both sets of hopping parameters
were constructed to produce similar undoped Fermi 
surfaces but their full band structures are different. 
However, it must be emphasized that  with both
sets of hoppings, charge stripes are found in several cases 
and, thus, their existence is a qualitative conclusion 
of our HF investigations.

\begin{figure}[thbp]
\begin{center}
\includegraphics[width=8.5cm,clip,angle=0]{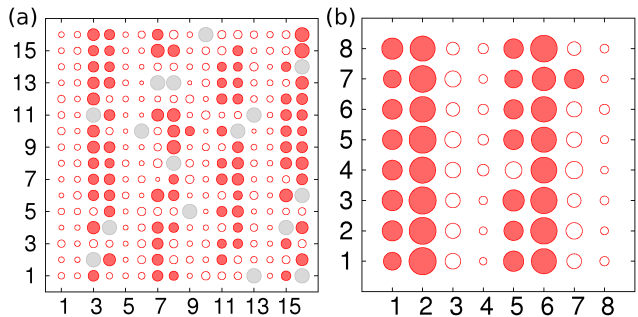}
\caption{(color online)  
(a) Ground state 
of the HF Hamiltonian at $\langle n \rangle$=2.56, $U$=1.0, and $J_{\rm H}$/$U$=0.25,
using the SK hoppings.\cite{maria,moreo} 
In the 16 sites shown in grey, there is an on-site energy $\epsilon$=-0.85,
that simulates the presence of quenched disorder, such as caused by Co doping.
(b) Monte Carlo results (equilibrated snapshot) 
obtained using the two-orbital spin-fermion model 
of Ref.~\onlinecite{SF} 
with SK hoppings,\cite{maria,moreo} $\langle n \rangle$=2.55, 
$K$=-1, $J_{\rm NN}$=$J_{\rm NNN}$=0.05, magnitude
of the classical spins $S$=1, temperature 0.005 eV, and 45,000 steps. The small
deviation from a perfect stripe is caused by temperature.
}
%\vskip -0.6cm
\label{Fig4}
\end{center}
\end{figure}

\subsection{Preliminary studies of the influence of Co doping on a striped state}

The striped states found here could be of relevance 
to explain the static nematic
state found via STM.\cite{seamus} To test this hypothesis, 
a randomly-chosen set of
lattice sites was selected to simulate the presence 
of Co-dopants via an on-site
energy $\epsilon$, and then the minimization of the HF model 
was again carried out
in that background. A typical result is shown in Fig.~\ref{Fig4}(a). It is 
observed that this quenched disorder effectively cuts the long stripes 
into shorter segments, and the overall state 
qualitatively resembles those found 
with STM. Of course, only a tuning of couplings 
and electronic densities may render
this agreement quantitative (for instance producing features of size 8 lattice
spacings, as in experiments) 
and that effort is postponed for future work. Here, simply note that the
charge and spin stripes found in our study tend to be perpendicular to one 
another, in nice agreement with the STM investigations.

\subsection{Results for the spin fermion model}

Striped states were also studied via the two-orbital
spin-fermion model recently proposed.\cite{SF} Using 
Monte Carlo techniques,\cite{manganites}
metastabilities limited the present effort to 8$\times$8 clusters. In cases where
good convergence was achieved, stripes were also found (Fig.~\ref{Fig4}(b)).
Thus, via the use of two rather different models and techniques the presence
of charge non-uniformed states was confirmed.

%{\it Conclusions.}

\section{Conclusions}
 
In summary, the numerical 
solution of the two-orbital Hubbard model in the HF approximation 
away from half-filling has been here discussed. Charge stripes appear at several
electronic densities and in a broad range of couplings. The associated spin and orbital properties of the
HF states have been discussed as well. However, the readers should be warned 
that the relevance of these states to the
real pnictides is a matter that still requires further work. 
On one hand, the charge stripes found here appear to have an associated spin incomensurability. 
Since neutron scattering results have actually unveiled a similar
spin incommensurability in cases where nesting can be excluded, then it is important to search for
alternative explanations such 
as those presented here. In addition, the stripes discussed in this effort 
can also provide
a qualitative rationalization of the nematic order found in STM experiments.
On the other hand, the HF striped states are the more stable at couplings $U$ where the
undoped state is an insulator, albeit with a small gap. However, for
the hopping sets investigated here, novel (weak) stripes were observed in the
magnetic/metallic region, thus avoiding this 
conceptual problem. But this interesting observation must
be confirmed with more refined calculations and using models with more orbitals. 

In addition, it may occur that 
the surface of pnictides presents an effective $U$ larger than
in the bulk, due to the reduction of the carrier's bandwidth, complicating the analysis. 
Another issue to consider is that the use of models with more than two orbitals
may lead to an effective electronic population in the $d_{xz}$-$d_{yz}$ sector that is larger 
than 2.0, and charge stripes, magnetism, and metallicity may coexist in a single state
at couplings $U$ larger than those estimated to be realistic 
from the $\langle n \rangle$=2.0 two-orbital model analysis. All these scenarios are
speculative right now, and only further work can clarify the relevance of the
striped states  presented here to the physics of the pnictides.

\section{Acknowledgments} 

This work was supported by 
the U.S. Department of Energy, Office of Basic Energy Sciences,
Materials Sciences and Engineering Division (Q.L., A.M., E.D.), and
the Sun Yat-sen University Hundred Talents program and NSFC-11074310 (D.X.Y.).
The computational studies were performed using the Kraken supercomputer
of the National Institute for Computational
Sciences and the Newton HPC Cluster 
at the University of Tennessee. The help of G. Alvarez 
and R. Yu for the programming of the
spin-fermion and HF models is acknowledged.

\end{document}